\documentclass[aps,prb,reprint,superscriptaddress,longbibliography,nofootinbib,floatfix]{revtex4-2}
\usepackage{amsmath,amssymb,bm,graphicx}
\usepackage[colorlinks=true,linkcolor=blue,citecolor=blue,urlcolor=blue]{hyperref}

\providecommand{\email}[1]{}
\makeatletter
\@ifundefined{aps@abstractbox}{}{%
  \gdef\@affiliation{}%
  \renewcommand{\affiliation}[1]{%
    \ifx\@affiliation\@empty
      \gdef\@affiliation{#1}%
    \else
      \g@addto@macro\@affiliation{\\#1}%
    \fi
  }%
}
\makeatother

\newcommand{\dd}{\mathrm d}
\newcommand{\ii}{\mathrm i}
\newcommand{\DeltaZ}{\Delta_0}
\newcommand{\xiz}{\xi_D}
\newcommand{\F}{\mathcal F}
\newcommand{\G}{\mathcal G}
\newcommand{\Lop}{\mathcal L}

\newcommand{\Bs}{B_{\rm s}}

\begin{document}

\title{Microscopic theory of the field-induced instability of the vortex-free state in superconducting thin-film strips}

\author{Takayuki Kubo}
\email{kubotaka@post.kek.jp}
\affiliation{High Energy Accelerator Research Organization (KEK), Tsukuba, Ibaraki 305-0801, Japan}
\affiliation{The Graduate University for Advanced Studies, SOKENDAI, Hayama, Kanagawa 240-0193, Japan}

\begin{abstract}
In the Pearl--London theory, the edge-barrier-disappearance field of a superconducting thin-film strip depends on an arbitrary short-distance core cutoff because the vortex is treated as a point object.  
The theory does not determine the cutoff or how it depends on temperature $T$, and therefore cannot determine the $T$ dependence of the instability field.  
Here we formulate the microscopic stability problem directly for the vortex-free superconducting state.  
This removes the core-cutoff ambiguity and determines the instability field $\Bs$ over the full temperature range and across all width regimes considered here.
For a homogeneous dirty strip with negligible self-field, three width regimes occur.  
For $W<W_1(T)$, superconductivity disappears continuously into the normal state through a one-dimensional (1D)
instability.  
For $W_1(T)<W<W_2(T)$, an edge-selective two-dimensional (2D) long-wavelength mode becomes unstable.  
For $W>W_2(T)$, the critical wave number is finite and the unstable mode is localized near an edge.  
In the wide-strip limit, $B_s\propto1/W$, recovering the Pearl--London scaling.  
In sufficiently narrow strips, however, the Pearl--London edge-barrier picture fails qualitatively.
\end{abstract}

\maketitle


\paragraph*{\bf Introduction.}

\begin{figure}[t]
\includegraphics[width=\columnwidth]{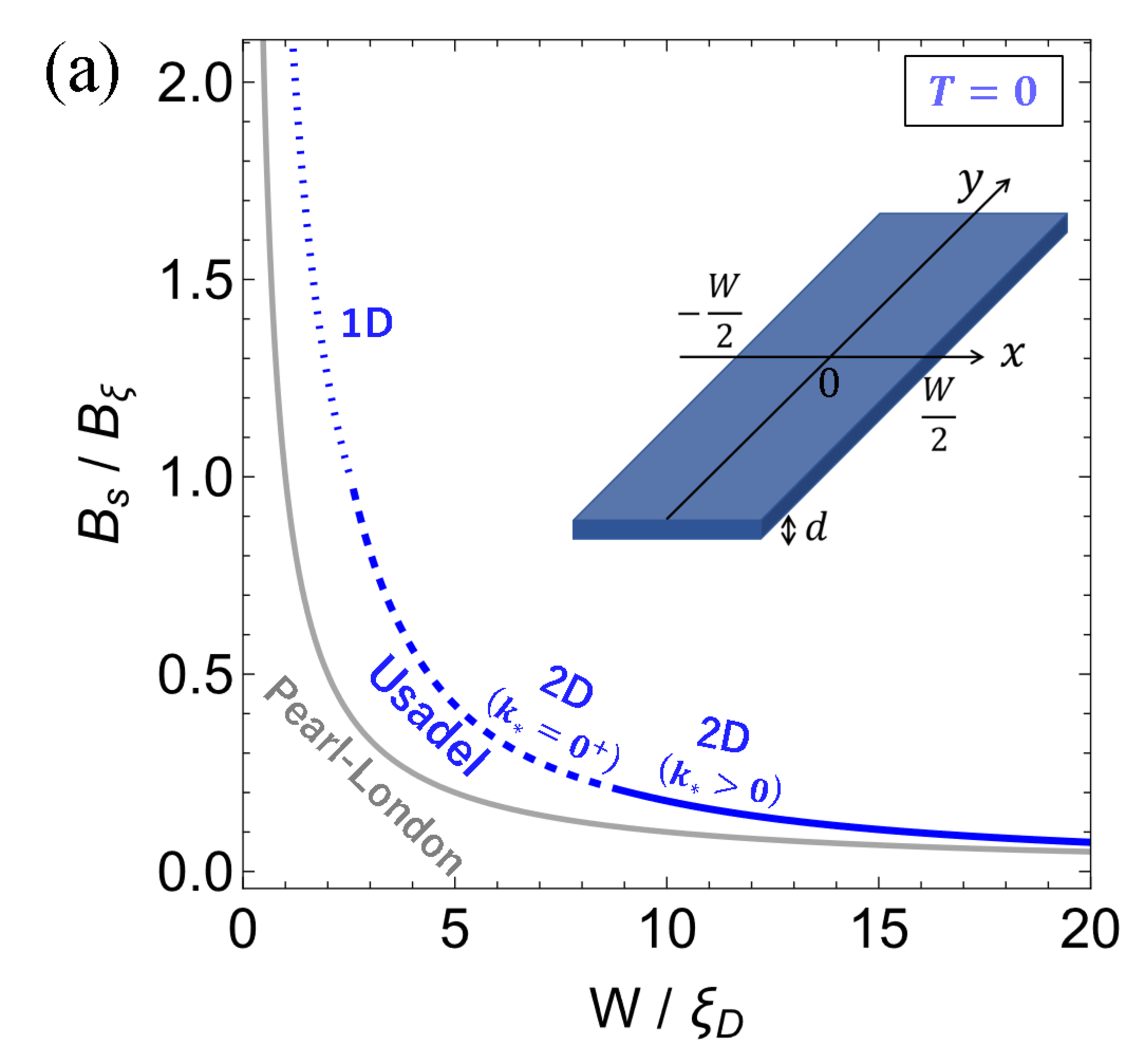}
\includegraphics[width=\columnwidth]{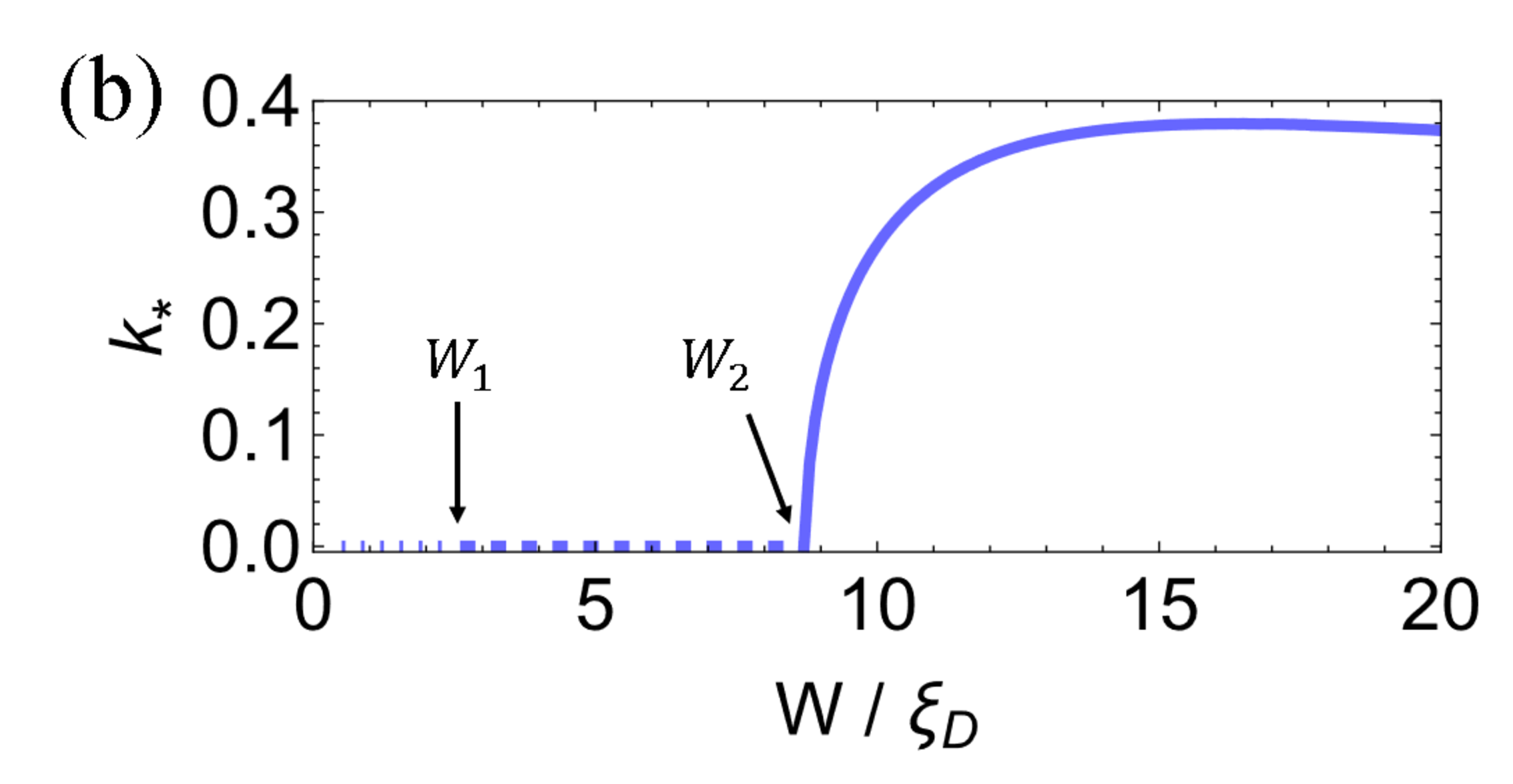}
\caption{
(a) The instability field $B_s$ as a function of the strip width $W$
at $T=0$.  The dotted, dashed, and solid segments indicate the
regimes in which $B_s$ is determined by the 1D, 2D long-wavelength,
and finite-wave-number instabilities, respectively.  The gray curve
shows the naive Pearl--London result
$B_s^{\rm (L)}/B_{\xi}=1/w$, obtained by setting
$\xi_{\rm cut}=\xi_D$.  The inset shows the strip geometry.
(b) Critical longitudinal wave number in the 2D sector.  The solid
segment shows the finite-wave-number regime with $k_*>0$. The dashed
segment at $k_*=0$ denotes the long-wavelength limit $k\to0^+$ with
$u_0\neq0$, whereas the dotted segment at zero indicates the 1D
regime, in which no 2D wave number determines $B_s$.  
}
\label{fig1}
\end{figure}

A superconducting thin-film strip in a perpendicular magnetic field can remain in a vortex-free state even when the applied field exceeds the lower critical field.  
In the Pearl--London theory~\cite{Likharev,Martinis,Kuit,Kogan_2020,Kogan_2021,Maksimova,Kubo_2023}, 
this metastability is described by an edge barrier that prevents a point
vortex from entering the strip.  
The barrier is obtained from the London Gibbs energy $G(x_v;B)$ as a function of the transverse vortex
position $x_v$.
Here, the strip extends along the $y$ direction and occupies $-W/2<x<W/2$ [see the inset of Figure~\ref{fig1} (a)].  
The Pearl length is $\Lambda=2\lambda^2/d$, where $\lambda$ is the London penetration depth and $d$ is the film thickness.  
In the narrow-strip limit $W\ll\Lambda$, the magnetic self-field can be neglected, and the applied field penetrates the film almost uniformly.

We denote the Pearl--London edge-barrier-disappearance field by $\Bs^{\rm (L)}$.  
It is defined as the applied field at which the barrier in $G(x_v;B)$ disappears and is given by
$\Bs^{\rm (L)}=\phi_0/(2\pi W\xi_{\rm cut})$~\cite{Maksimova,Kubo_2023}, where $\phi_0$ is the superconducting flux quantum and $\xi_{\rm cut}$ is a short-distance cutoff of the order of
the coherence length.  
This construction is physically transparent and widely useful, but it is not a microscopic stability calculation.
The vortex is treated as a singular point object, and the value of $\xi_{\rm cut}$ must therefore be supplied separately.  
The Pearl--London theory does not determine this cutoff or how it depends on temperature, so the temperature dependence of $\Bs^{\rm (L)}$ remains ambiguous.

A microscopic treatment that removes this ambiguity should not begin by placing a London point vortex at a prescribed position.  
Instead, it should examine the stability~\cite{Strogatz} of the vortex-free superconducting state.  
At low applied fields, the vortex-free solution is a local minimum of the Gibbs functional.  
If its second variation is positive for every allowed physical perturbation, any path away from this state
must begin by increasing the energy.  
As the applied field increases, the lowest physical curvature decreases and eventually vanishes.  
We define $\Bs$ as the field at which it first reaches zero.  
Thus, $\Bs$ is the spinodal field of the vortex-free state and the microscopic counterpart of the Pearl--London barrier-disappearance field.  
In this sense, $\Bs$ is the thin-film-strip counterpart of the superheating field of a semi-infinite bulk superconductor~\cite{Catelani,Transtrum,Gurevich_2023,Kubo_2026}.

This Letter formulates the field-induced instability of a homogeneous dirty thin-film strip within the Usadel theory.  
We first examine one-dimensional (1D) perturbations that depend only on the transverse coordinate and then generalize the stability problem to two-dimensional (2D) perturbations with a longitudinal Fourier wave number, including the long-wavelength limit.  
The formulation removes the adjustable core-cutoff ambiguity and applies over the full temperature range below $T_c$.

This work completes a three-part microscopic description of the basic vortex-related thresholds in narrow superconducting thin-film strips.
Together with the microscopic theory of the lower critical field, which removes the core-cutoff ambiguity~\cite{Kubo_Bc1}, and the proof that the barrier-disappearance current equals the depairing current in
an ideal strip~\cite{Kubo_Jv}, the present theory of the field-induced instability provides a unified microscopic description of these thresholds.

\paragraph*{\bf Usadel functional.}

We consider a homogeneous dirty superconducting strip [see the inset of Fig.~\ref{fig1} (a)] in a static perpendicular magnetic field \(B\hat{\bf z}\), with no transport current.  
The strip is infinite in the \(y\) direction and has width \(W\) in the \(x\) direction. 
The film is thin, \(d\ll\lambda\), and we take \(W\ll\Lambda=2\lambda^2/d\), 
so that the magnetic self-field of the induced sheet current is neglected.  
The applied vector potential is fixed externally.  
We use the gauge \({\bf A}=xB\hat{\bf y}\).
We follow the convention of Refs.~\cite{Gurevich_Kubo, Kubo_2020, Kubo_erratum}.  
The gauge-invariant momentum is \({\bf q}=\nabla\chi+(2\pi/\phi_0){\bf A}\); with this sign convention, the supercurrent flows opposite to \({\bf q}\).  
The Matsubara Green functions are parametrized as \(g_n=\cos\theta_n\) and \(f_n=\sin\theta_n e^{\ii\chi}\).  
The magnetic field is measured in units of $B_\xi:=\phi_0/(2\pi\xiz^2)$, 
where $\xiz=\sqrt{\hbar D/(2\DeltaZ)}$ is the temperature-independent dirty-limit coherence length, $D$ is the diffusion constant, and $\DeltaZ$ is the zero-temperature gap in the absence of pairbreaking.
We write $b:=B/B_\xi$.
Lengths are measured in units of \(\xiz\), and we introduce the dimensionless variables \(\bar{\nabla}:=\xiz\nabla\), \(\bar{\bf r}:={\bf r}/\xiz=(\bar{x},\bar{y})^T\), \(w:=W/\xiz\), and \({\bf Q}:=\xiz{\bf q}=\bar{\nabla}\chi+b\bar{x}\hat{\bf y}\).  
The gap and Matsubara energies are measured in units of $\DeltaZ$;
accordingly, $\bar{\Delta}:=\Delta/\DeltaZ$ and $\Omega_n:=\hbar\omega_n/\DeltaZ=2\pi\tau(n+1/2)$, where $\tau=k_BT/\DeltaZ$.  
In what follows, we omit the bars for brevity.
An overall positive prefactor of the free-energy functional is also omitted, since it does not affect the stationary equations or the stability condition.

In equilibrium, the system is governed by
\begin{eqnarray}
&&\nabla^2\theta_n = |{\bf Q}|^2\sin\theta_n\cos\theta_n
+\Omega_n\sin\theta_n - \Delta\cos\theta_n,
\label{eq:usadel} \\
&&\Delta\ln\frac{T}{T_c}
+2\pi\tau\sum_{n=0}^{\infty}
\left(\frac{\Delta}{\Omega_n}-\sin\theta_n\right)=0,
\label{eq:gap} \\
&&\nabla\cdot[S({\bf r}){\bf Q}({\bf r})]=0,\qquad
S({\bf r})=2\pi\tau\sum_{n=0}^{\infty}\sin^2\theta_n({\bf r}).
\label{eq:phase_equation}
\end{eqnarray}
These equations are, respectively, the Usadel equation, the gap equation, and current conservation.  
The current density is proportional to $-S{\bf Q}$.
They are obtained by imposing the stationarity conditions $\delta\F/\delta\theta_n=0$, $\delta\F/\delta\Delta=0$, and $\delta\F/\delta\chi=0$ on the Usadel free-energy functional, measured relative to the normal state,
\begin{eqnarray}
\F[\{ \theta_n\}, \Delta, \chi]
&=& \int \dd^2r\,
\biggl\{
\Delta^2\ln\frac{T}{T_c}
+2\pi\tau\sum_{n=0}^{\infty}
\biggl[ \frac{\Delta^2}{\Omega_n}
\nonumber\\
&+&|\nabla\theta_n|^2
-2\Delta\sin\theta_n
+|{\bf Q}|^2\sin^2\theta_n
\nonumber\\
&+& 2\Omega_n(1-\cos\theta_n)
\biggr]
\biggr\}.
\label{eq:free_energy}
\end{eqnarray}
Here the applied field enters through ${\bf Q}=\nabla\chi+b x\hat{\bf y}$ and is an externally prescribed,
non-variational parameter.  
The magnetic self-field is neglected throughout. 
At \(T=0\), the replacement \(2\pi\tau\sum_{n\ge0}\rightarrow \int_0^\infty\dd\Omega\) is made after the standard gap-equation regularization.

The vortex-free solution at fixed applied field is translation invariant along $y$.  
We take $\chi=0$, so that ${\bf Q}(x)=Q(x)\hat{\bf y}$ with $Q(x)=bx$.  
The gap and spectral angles depend only on $x$.  
The Usadel equation reduces to
\begin{equation}
\theta_n'' -Q^2(x)\sin\theta_n\cos\theta_n -\Omega_n\sin\theta_n +\Delta\cos\theta_n
=0,
\label{eq:background_usadel}
\end{equation}
together with Eq.~\eqref{eq:gap}.  
The insulating strip edges impose $\theta_n'(\pm w/2)=0$.  
The screening current is antisymmetric in $x$, and its integral across the strip vanishes; 
hence there is no net transport current.  
The local pair breaking is largest near the two edges because $|Q(x)|$ is largest there.

For a fixed applied field, stability is determined by the Gibbs free energy.  
Relative to the normal state in the same applied field, the Gibbs functional differs from $\F$ only by the magnetic energy associated with the self-field.  
Since the self-field is neglected in the present approximation, this magnetic contribution vanishes.
Therefore, $\G=\F$ for the present stability calculation.  
For sufficiently small $b$, the vortex-free solution is locally stable:
its quadratic variation is positive for all nonzero perturbations except for the trivial global-phase zero mode.
We define the instability field $B_s$ as the value of $B$ at which the lowest physical curvature first vanishes.  
Its dimensionless form is $b_s=B_s/B_\xi$.

\paragraph*{\bf 1D perturbations.}

We first consider strictly 1D perturbations:
\begin{eqnarray}
&& \chi\to\varphi(x), \nonumber\\
&& \Delta\to\Delta+\eta(x), \nonumber\\
&& \theta_n\to\theta_n+\alpha_{n}(x).
\label{eq:1d_perturbations}
\end{eqnarray}
Since the applied vector potential is fixed, the momentum perturbation is \(\delta{\bf Q}=\varphi'(x)\hat{\bf x}\).
The insulating-edge conditions are \(\varphi'(\pm w/2)=0\) and \(\alpha_{n}'(\pm w/2)=0\).
The quadratic variation is
\begin{eqnarray}
\delta^2\G_{\rm 1D}
&=&
\int_{-w/2}^{w/2}\dd x
\biggl\{
\eta^2\ln\frac{T}{T_c}
+2\pi\tau\sum_n
\biggl[
\frac{\eta^2}{\Omega_n}
\nonumber\\
&+&
(\alpha_{n}')^2
+d_n\alpha_{n}^2
-2c_n\eta\alpha_{n}
+
s_n^2(\varphi')^2
\biggr]
\biggr\},
\label{eq:quadratic_1d}
\end{eqnarray}
where \(s_n(x):=\sin\theta_n(x)\), \(c_n(x):=\cos\theta_n(x)\), and $d_n(x):=(c_n^2-s_n^2)Q^2+\Omega_n c_n+\Delta s_n$. 
The phase perturbation is decoupled from $\eta$ and $\alpha_n$ and contributes $\int\dd x\,S(x)(\varphi')^2\ge0$.  
Since $S(x)>0$ in the superconducting state, equality occurs only for a spatially uniform global-phase rotation.  
A 1D instability can therefore occur only in the coupled gap-amplitude and spectral-angle sector.

For a given \(\eta\), minimization with respect to \(\alpha_{n}\), namely $\delta_{\alpha_{n}} (\delta^2 \G_{\rm 1D})=0$, gives a linear equation relating $\alpha_n$ to $\eta$: $\Lop_n\alpha_n=c_n\eta$, where $\Lop_n:=-\dd^2/\dd x^2+d_n(x)$. 
Eliminating $\alpha_n$ therefore gives the reduced quadratic form 
\begin{eqnarray}
\delta^2\G_{\rm 1D}^{\rm red}
&=& \int\dd x\, \eta^2\ln\frac{T}{T_c}
+2\pi\tau\sum_n \int\dd x \biggl[ \frac{\eta^2}{\Omega_n} \nonumber\\ &&
- c_n\eta\, \Lop_n^{-1} \left(c_n\eta\right) \biggr].
\label{eq:reduced_1d}
\end{eqnarray}
As shown below, the 1D stability limit coincides with the field at which the superconducting solution continuously reaches the normal state.  
We therefore evaluate Eq.~\eqref{eq:reduced_1d} for $\Delta=\theta_n=0$.
In the normal state, $c_n=1$ and $\Lop_n=\Omega_n+\mathcal A_b$, where $\mathcal A_b:=-\dd^2/\dd x^2+b^2x^2$.
Let
\begin{eqnarray}
\mathcal A_b\varphi_j = a_j(b,w)\varphi_j , \label{eq:op_A}
\end{eqnarray}
subject to $\varphi_j'(\pm w/2)=0$.
We choose the eigenfunctions to be orthonormal, \(\int_{-w/2}^{w/2}\dd x\,\varphi_i(x)\varphi_j(x)=\delta_{ij}\), and expand the gap perturbation as \(\eta=\sum_j\eta_j\varphi_j\).  
Since \(\Lop_n^{-1}\varphi_j=\varphi_j/[\Omega_n+a_j(b,w)]\), Eq.~\eqref{eq:reduced_1d} becomes
\begin{eqnarray}
&&\delta^2\G_{\rm 1D, N}^{\rm red}\nonumber \\
&=&
\sum_j
\biggl[
\ln\frac{T}{T_c} +2\pi\tau\sum_{n=0}^{\infty}
\biggl(
\frac{1}{\Omega_n} 
-
\frac{1}{\Omega_n+a_j(b,w)}
\biggr)
\biggr]
|\eta_j|^2
\nonumber\\
&=&
\sum_j
\biggl[
\ln\frac{T}{T_c}
+\psi\left(
\frac{1}{2}
+\frac{a_j(b,w)}{2\pi\tau}
\right)
-\psi\left(\frac{1}{2}\right)
\biggr]
|\eta_j|^2 .
\label{eq:normal_reduced_1d}
\end{eqnarray}
The coefficient in square brackets increases monotonically with \(a_j\).  
The lowest curvature therefore corresponds to the lowest orbital eigenvalue \(a_0(b,w)\).  
The field at which this curvature vanishes is determined by
\begin{eqnarray}
\ln\frac{T}{T_c} + \psi\left( \frac{1}{2} +\frac{a_0(b_s^{\rm (1D)},w)}{2\pi\tau} \right)
- \psi\left(\frac{1}{2}\right) = 0 ,
\label{eq:usadel_1d_endpoint}
\end{eqnarray}
where $\psi$ is the digamma function.  
For a given temperature, Eq.~\eqref{eq:usadel_1d_endpoint} first determines the critical value
of $a_0$.  
The 1D instability field $b_s^{\rm (1D)}$ is then obtained by solving the eigenvalue problem in Eq.~\eqref{eq:op_A} and finding the value of $b$ for which its lowest eigenvalue $a_0(b,w)$ reaches this critical value.
At $T=0$, Eq.~\eqref{eq:usadel_1d_endpoint} is understood as the limit $T\to0$, 
which gives $a_0(b_s^{\rm (1D)},w)=1/2$.

We now show that superconductivity reaches the normal state continuously through the 1D instability at $b=b_s^{\rm (1D)}$.
 At this field, the zero-curvature perturbations introduced in Eq.~\eqref{eq:1d_perturbations} satisfy $(\Omega_n+\mathcal A_b)\alpha_n=\eta$.  
The quadratic contribution vanishes along this direction.  
Expanding the Gibbs functional to fourth order, using this linear relation, and integrating by parts gives $\G-\G_{\rm N}=2\pi\tau\sum_n\int\dd x\,[\alpha_n^2(\alpha_n')^2+\Omega_n\alpha_n^4/4]$, apart from terms of sixth and higher order.  
The fourth-order contribution is therefore positive. 
For $b$ slightly below $b_s^{\rm (1D)}$, the lowest eigenvalue $a_0(b,w)$ is smaller than the critical value determined by Eq.~\eqref{eq:usadel_1d_endpoint}.  
The quadratic coefficient along the corresponding mode is therefore negative, 
so that a small superconducting order parameter lowers the Gibbs energy relative to the normal state.  
As $b$ increases toward $b_s^{\rm (1D)}$, $a_0(b,w)$ approaches its critical value and the quadratic coefficient
increases continuously to zero.  
Together with the positive fourth-order contribution, this means that the minimum remains at a nonzero order-parameter amplitude below $b_s^{\rm (1D)}$, but this amplitude continuously decreases to zero as the instability field is approached.  
The lowest 1D curvature also vanishes continuously at $b=b_s^{\rm (1D)}$.
At finite temperatures, we have also verified numerically from Eq.~\eqref{eq:reduced_1d} that no other 1D curvature changes sign at a lower field over the temperature and width ranges studied.

The physical interpretation is as follows.  
The 1D perturbation is uniform along the entire length of the strip and does not select either an edge or a particular longitudinal position.  
Its transverse profile is given by the node-free lowest eigenfunction $\varphi_0(x)$.  
For a sufficiently narrow strip, this eigenfunction is nearly constant across the width, $\varphi_0(x)\simeq{\rm const.}$, so that the superconducting state is suppressed almost uniformly over
the entire strip.  
As the magnetic field increases, the overall amplitude of this profile decreases continuously to zero, and
superconductivity therefore disappears throughout the strip at $b=b_s^{\rm (1D)}$.

\paragraph*{\bf 2D perturbations.}

We next examine 2D perturbations that vary along the strip.  
Since the vortex-free background is translationally invariant along $y$,
different longitudinal Fourier components decouple.  
Modes with $k$ and $-k$ are equivalent, so it is sufficient to consider $k>0$.  
We write
\begin{eqnarray}
&& \chi \to \varphi_k(x)\frac{1}{\sqrt{\pi}}\sin ky,
\nonumber\\
&& \Delta \to \Delta+\eta_k(x)\frac{1}{\sqrt{\pi}}\cos ky,
\nonumber\\
&& \theta_n \to \theta_n+\alpha_{nk}(x)\frac{1}{\sqrt{\pi}}\cos ky .
\label{eq:2d_perturbations}
\end{eqnarray}
The factors $1/\sqrt{\pi}$ give the continuum normalization on the infinite $y$ axis.  
The equivalent perturbations obtained by interchanging the sine and cosine functions have the same eigenvalues and therefore do not need to be considered separately.
The corresponding momentum perturbation is  $\delta{\bf Q}_k = \nabla\left[\varphi_k(x)\sin ky\right]/\sqrt{\pi} = \varphi_k'(x)\sin ky\,\hat{\bf x}/\sqrt{\pi}+k\varphi_k(x)\cos ky\,\hat{\bf y}/\sqrt{\pi}$.
The boundary conditions are \(\varphi_k'(\pm w/2)=0\) and \(\alpha_{nk}'(\pm w/2)=0\).

It is useful to introduce $u_k(x):=k\varphi_k(x)$, because this quantity directly gives the longitudinal component of the momentum perturbation: $\delta Q_y=u_k(x)\cos ky/\sqrt{\pi}$.  
The transverse component is $\delta Q_x=u_k'(x)\sin ky/(k\sqrt{\pi})$.
After integration over $y$, the quadratic variation is
\begin{eqnarray}
\delta^2\G_{{\rm 2D},k}
&=&
\int_{-w/2}^{w/2}\dd x
\biggl\{
\eta_k^2\ln\frac{T}{T_c}
+2\pi\tau\sum_n
\biggl[
\frac{\eta_k^2}{\Omega_n}
\nonumber\\
&+&
(\alpha_{nk}')^2
+(d_n+k^2)\alpha_{nk}^2
-2c_n\eta_k\alpha_{nk}
\nonumber\\
&+&
4Qs_nc_nu_k\alpha_{nk}
+s_n^2
\left\{
u_k^2+\frac{(u_k')^2}{k^2}
\right\}
\biggr]
\biggr\}.
\label{eq:quadratic_local}
\end{eqnarray}
The last term is the direct quadratic energy cost associated with the momentum perturbation $\delta{\bf Q}$.  
To prevent this contribution from diverging as $k\to0^+$, one must have $u_k'=O(k)$, 
so that $u_k(x)$ approaches an $x$-independent constant $u_0$.

The limit $k\to0^+$ then contains two physically different cases.
If $u_0\neq0$, then $\varphi_k\simeq u_0/k$ and, over distances much shorter than $1/k$, the limiting perturbation is $\delta\chi\simeq u_0y/\sqrt{\pi}$ and $\delta Q_y\simeq u_0/\sqrt{\pi}$.  
The background momentum therefore changes as $Q_y(x)=bx\to bx+u_0/\sqrt{\pi}$.  
This change shifts the background momentum profile toward one edge, increasing the pair-breaking momentum near that edge and decreasing it near the opposite edge.
The mode thus distinguishes the two edges while remaining coherent along the entire strip.  We refer to this case as the 2D long-wavelength mode. 
If $u_0=0$, the additional longitudinal-momentum degree of freedom disappears, and the remaining amplitude and spectral-angle perturbations reduce to those of the 1D problem discussed above.

For every finite $k$, the longitudinal momentum perturbation has zero average along $y$.  
Thus, $k=0^+$ denotes the limit of this zero-average family and not an externally imposed uniform transport
current.

For given $\eta_k$ and $u_k$, minimization with respect to $\alpha_{nk}$ gives $\Lop_{nk}\alpha_{nk}=c_n\eta_k-R_nu_k$, where $\Lop_{nk}:=-\dd^2/\dd x^2+k^2+d_n(x)$ and $R_n(x):=2Q(x)s_n(x)c_n(x)$.  Substituting this solution into Eq.~\eqref{eq:quadratic_local} gives
\begin{eqnarray}
\delta^2\G_{{\rm 2D},k}^{\rm red}
&=&
\int_{-w/2}^{w/2}\dd x
\biggl\{
\eta_k^2\ln\frac{T}{T_c}
+2\pi\tau\sum_n
\biggl[
\frac{\eta_k^2}{\Omega_n}
\nonumber\\
&&
+s_n^2
\biggl\{
u_k^2+\frac{(u_k')^2}{k^2}
\biggr\}
\nonumber\\
&-&
\left(c_n\eta_k-R_nu_k\right)
\Lop_{nk}^{-1}
\left(c_n\eta_k-R_nu_k\right)
\biggr]
\biggr\}.
\label{eq:quadratic_2d_reduced}
\end{eqnarray}
This is a real symmetric quadratic form in $\eta_k$ and $u_k$.  Let
$\lambda_{\rm 2D}(b,k)$ denote its lowest nontrivial curvature, where
$k=0^+$ means the limit $k\to0^+$ at fixed
$u_k=k\varphi_k$.  The 2D instability field
$b_s^{\rm (2D)}$ and the critical wave number $k_*$ are defined by
\begin{equation}
\lambda_{\rm 2D}\!\left(b_s^{\rm (2D)},k_*\right)=0,
\label{eq:bs_2d}
\end{equation}
where $k_*$ gives the lowest such field among $k=0^+$ and all
$k>0$.  Only a zero reached while the background remains
superconducting is counted; the zero curvature of the normal state is
excluded.  The $k=0^+$ solution with $u_0=0$ reduces to the 1D
perturbation and is not counted separately.

\paragraph*{\bf Instability field.}

The instability field is given by
\begin{equation}
b_s
=
\min
\left\{
b_s^{\rm (1D)},
b_s^{\rm (2D)}
\right\} ,
\label{eq:Bs_final_definition}
\end{equation}
when a nontrivial 2D threshold exists; otherwise, $b_s=b_s^{\rm (1D)}$.

For the numerical results shown below, the spatial mesh, Matsubara-frequency cutoff, longitudinal-wave-number grid, and field resolution were refined independently.  
The estimated relative uncertainty in $B_s$ is below $0.05\%$.  
The numerical implementation and source code are provided separately.

\begin{figure}[t]
\includegraphics[width=\columnwidth]{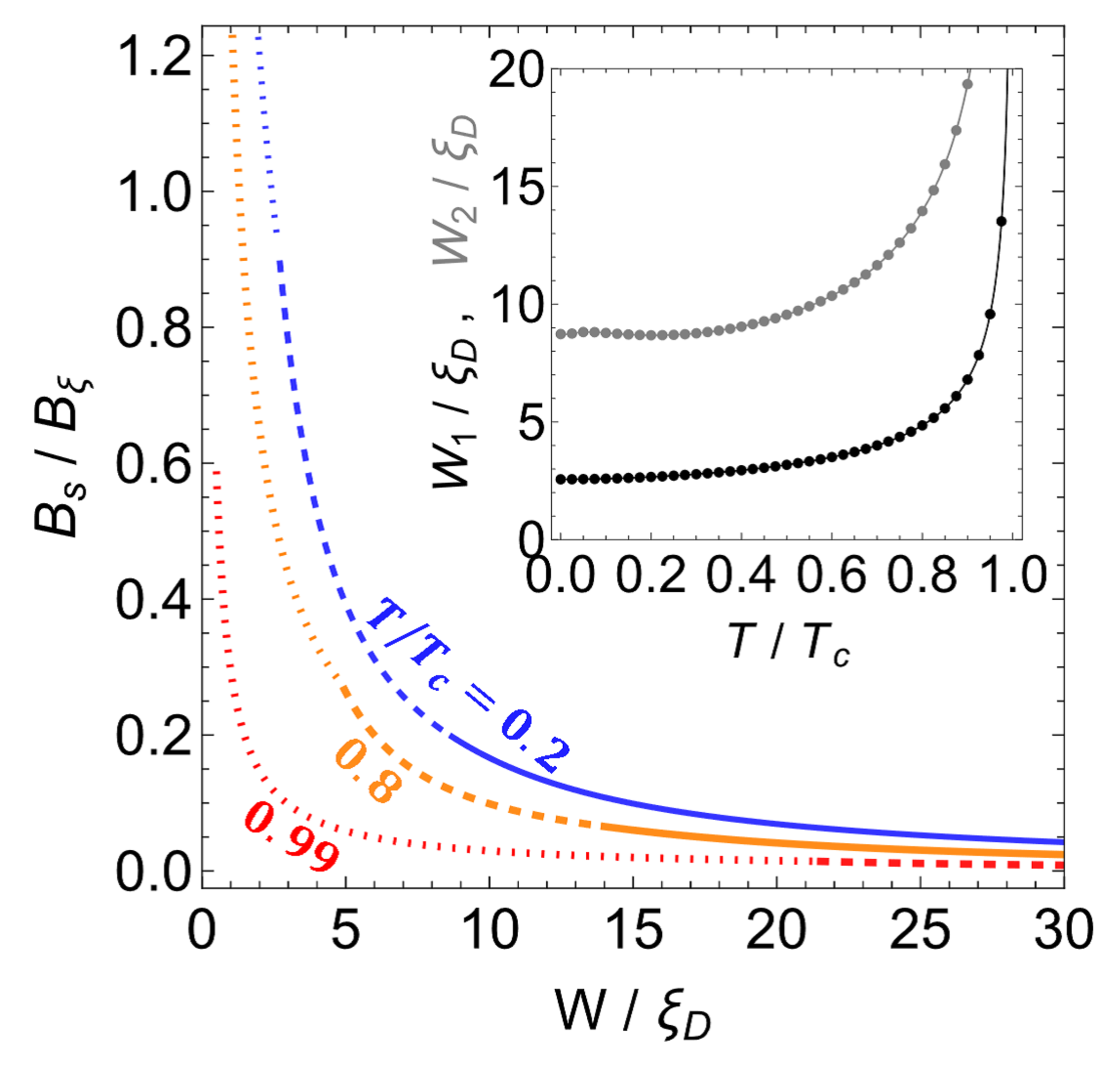}
\caption{
The instability field $B_s$ as a function of the strip width $W$ for
$T/T_c=0.2$, $0.8$, and $0.99$.  The dotted, dashed, and solid
segments indicate the regimes in which $B_s$ is determined by the 1D,
2D long-wavelength, and 2D finite-wave-number instabilities,
respectively.  The inset shows the temperature dependence of the
crossover widths $w_1=W_1/\xi_D$ and $w_2=W_2/\xi_D$.  The black and
gray curves are fits to $w_1$ and $w_2$, respectively:
$w_1(t)=(2.563+4.63t-0.125t^2)/
[(1+2.314t)\sqrt{1-t}]$ and
$w_2(t)=P[7t/(1+7t)]/\sqrt{1-t}$, where
$P(x)=8.73-1.45x+10.60x^2-37.20x^3+42.35x^4-18.46x^5$ and
$t=T/T_c$. Both fits reproduce the numerical results with a relative error of
about $0.05\%$ or less.
}
\label{fig2}
\end{figure}

Figures~\ref{fig1}(a) and \ref{fig1}(b) show the width dependence of $b_s$ and the corresponding critical wave number $k_*$ at $T=0$.
For $W<W_1$, $b_s=b_s^{\rm (1D)}$, and the order parameter decreases continuously to zero throughout the strip.  
This non-edge-selective instability cannot be described within the Pearl--London theory, which assumes a pre-existing point vortex and treats the loss of metastability as an edge-entry process.  
For $W_1<W<W_2$, an edge-selective 2D mode with $k_*=0^+$ becomes unstable while remaining coherent along the strip.  
At $W=W_2$, $k_*$ begins to increase continuously from zero; for $W>W_2$, the mode acquires a finite longitudinal length scale.

The behavior in a wide strip can be understood as follows.  
When $\xi_D\ll W\ll\Lambda$, $Q(x)=bx$ changes only slightly over a coherence-length-scale region near either edge.  
The vortex-free state in this edge region is therefore well approximated by a uniform current-carrying state.  
At the instability, the edge momentum is $Q_{\rm edge}=b_sw/2$.  
The instability occurs when this momentum reaches the depairing momentum $Q_d(T)$, 
giving $b_s\simeq2Q_d(T)/w$ for $w\gg1$.
At $T=0$, the uniform dirty-limit Usadel solution gives $Q_d(0)=[\zeta_d\exp(-\pi\zeta_d/4)]^{1/2}=0.4871$, where $\zeta_d=[16+3\pi^2-\sqrt{256+32\pi^2+9\pi^4}]/(8\pi)=0.3004$~\cite{Kubo_2020,Kubo_erratum,Kubo_2021, Kubo_2025}; see also Refs.~\cite{Maki,Kupriyanov,Clem_Kogan}.  
Consequently, $b_s\simeq0.9742/w$.  
The naive Pearl--London expression $b_s^{\rm (L)}=1/w$, obtained by setting $\xi_{\rm cut}=\xi_D$, reproduces the correct $1/w$ scaling and differs from the microscopic asymptotic coefficient by only about $2.6\%$.  
Finite-width corrections remain visible over the range shown in Fig.~\ref{fig1}(a).

Figure~\ref{fig2} shows the width dependence of $B_s$ at finite temperatures.  
Since $\xi_D$ is temperature independent, the dimensionless width $w=W/\xi_D$ of a given strip is fixed. 
Both crossover widths increase with temperature because the characteristic length scale of the superconducting response grows as $T$ approaches $T_c$.  
Near $T_c$, this increase can be attributed to the divergence of $\xi_{\rm GL}(T)$.
The inset shows $w_1=W_1/\xi_D$ and $w_2=W_2/\xi_D$.  
Near $T_c$, the numerical fits approach $2.1328/\sqrt{1-t}$ and $6.0123/\sqrt{1-t}$, in close agreement with the GL results $2.1331/\sqrt{1-t}$ and $6.0090/\sqrt{1-t}$, respectively; 
see the Supplemental Material for the GL calculation.

The weak nonmonotonicity of $W_2(T)$ reflects a competition between the energy cost of forming a modulation along the edge and the energy gain obtained by jointly changing the gap amplitude and the superfluid momentum.  
Since these two contributions have different temperature dependences, their balance changes at low temperatures.  
The energy cost is relatively stronger near $T/T_c\simeq0.06$, 
whereas the energy gain becomes more important at intermediate temperatures.  Above
$T/T_c\simeq0.21$, the growth of the coherence length drives $W_2$ upward again.

\paragraph*{\bf Discussion and conclusion.}

We have developed a microscopic theory of the field-induced instability of a homogeneous dirty thin-film strip.  The formulation removes the core-cutoff ambiguity of the Pearl--London treatment and determines $B_s$ over the full temperature range below $T_c$.  
Three width regimes occur: a continuous disappearance of superconductivity through a 1D instability, an edge-selective 2D long-wavelength instability, and a finite-wave-number 2D instability.  
In the large-$w$ limit at $T=0$, $b_s\simeq0.9742/w$, reproducing the Pearl--London $1/w$ scaling within $2.6\%$.  In sufficiently narrow strips, however, the edge-barrier picture fails qualitatively.

The calculation determines $B_s$ as the field at which the vortex-free state ceases to be a local minimum of the Gibbs functional.  
The edge-selective 2D modes strongly suggest vortex nucleation and entry, but establishing what happens beyond the instability requires a separate nonlinear calculation~\cite{Vodolazov_2012}.  
The present calculation assumes a homogeneous strip with no transport current and negligible self-field; extensions beyond these assumptions remain subjects for future work.

\begin{acknowledgments}
This work was supported by JSPS KAKENHI under Grant Nos. JP25K01610, JP25K23386, JP26K03209, and JP26K00665. 
The idea for this work emerged during my three-year paternity leave from 2021 to 2024. I am deeply grateful to everyone who supported me during that period, which was made possible by the Act on Childcare Leave of Japan~\cite{ikuji}.
\end{acknowledgments}

\end{document}